\documentclass[fleqn,twoside]{article}
\usepackage{gc,epsf}
\usepackage{graphics}
\heads{S.M.Kozyrev}
      {''JORDAN'S SCALAR STARS''AND DARK MATTER.}

\begin{document}
\twocolumn[

\Title{''JORDAN'S SCALAR STARS''AND DARK MATTER.}

\Author{S.M.KOZYREV\foom 1}
       {Scientific center gravity wave studies ''Dulkyn'',
Kazan, Russian Federation}

\Abstract
    {Here we are starting the study of the field
equations of relativistic scalar tensor theories in the
spherically symmetric gravitational field. In the present article
we shall consider as an example only the simplest
Jordan-Brans-Dicke (JBD) one. To illustrate the property of the
spherically symmetric JBD configuration we exhibit a new
representation of the well-known four dimensional solutions. In
this model, a suitable segment of Brans solution is chosen for the
interior of the object while the outer region consists of a
Schwarzschild vacuum. We have constructed "Jordan's scalar stars"
model consisting of three parts: a homogeneous inner core with
equation of state \textit{p}$_M$ = $\varepsilon $ $\varrho _M$; an
envelope of Brans spacetime matching the core and the exterior
Schwarzschild spacetime. We have also showed that this toy model
can explain the intergalactic effects without the dark matter
hypothesis.}

%\RAbstract  {"ЙОРДАНОВСКИЕ СКАЛЯРНЫЕ ЗВЕЗДЫ" И ТЕМНАЯ МАТЕРИЯ} {С.
%М. Козырев} {Мы начинаем исследование сферически-симметричных
%конфигураций в релятивистских скалярно-тензорных теориях. В данной
%статье мы рассматриваем только самый простой случай теорию
%Йордана-Бранса-Дикке. Чтобы проиллюстрировать свойства
%сферически-симметричных JBD конфигураций, мы предлагаем новую
%интерпретацию известных четырехмерных решений. В этой модели
%решения Бранса рассматриваются как внутренняя область объекта,
%которая сшивается с вакуумным решением Шварцшильда. Нами получены
%решения уравнений поля, которые описывают "Йордановские скалярные
%звезды" с сердцевиной из несжимаемой жидкости, скалярной оболочкой
%и окруженные Шварцшильдовским вакуумом. Мы показали, что эта
%грубая модель может объяснить межгалактические гравитационные
%эффекты без привлечения гипотезы темной материи.}

]

\email 1 {Sergey@tnpko.ru}

\section{Introduction}

The spirit of scalar-tensor extension of general relativity is an
attempt to properly incorporate the Mach's principle \cite{Mach}
and Dirac's large number hypothesis \cite{Dirac} in which Newton's
constant is allowed to vary with space and time. Apart from this,
it is known that the Jordan-Brans-Dicke (JBD) scalar field plays
the role of classical exotic matter required for the construction
of traversable Lorentzian wormholes \cite{Visser}, \cite{Nandi}.
The most prominent example of scalar-tensor theories of gravity is
perhaps the
 JBD theory \cite{Jordan}, \cite{Brans}. These theories
introduce a new fundamental scalar field which appears to be coupled
non-minimally to gravity (in the so-called Jordan frame).

It is usually believed that when the effective JBD parameter
$\omega $ is sufficiently large, the scalar-tensor theories of
gravity are compatible with the solar system tests. However, a
number of exact JBD solutions have been reported not to tend to
the corresponding general relativity solutions \cite{Matsuda},
\cite{Romero},\thinspace \cite{Scheel}. These situation are
alarming since the standard belief that JBD theory always reduces
to general relativity in the large $\omega $ limit is the basis
for setting lower limits on the $\omega $-parameter using
celestial mechanics experiments \cite {Will},
\cite{Fujii},\thinspace \cite{Eubanks}. To make the situation
worse, as showed by Hawking \cite{Hawking} and Johnson
\cite{Johnson} (cf. Thorne and Dykla \cite{Thorne}) the only black
holes in the JBD theory are Einstein black holes. If, following
Hawking theorem \cite{Hawking}, we make the reasonable demand that
the solution of scalar-tensor theory field equations in empty
space is the Schwarzschild solution lead to free estimates of
lower limit of $\omega $. Moreover, when the energy momentum
tensor of ordinary matter vanishes, for all values of $\omega $
the JBD theory can agrees with Einstein theory up to any desired
accuracy and hence observations cannot rule out the JBD theory in
favor of general relativity. Note, it is known that in empty space
some vector-metric theories \cite
{Will2} can be recast simply in the Einstein's theory with $\stackrel{%
\rightarrow }{K}$= 0, but inside a matter the potential for the
effective vector degree of freedom may play a complicated and
nontrivial role  \cite{Bashkov}. It is thus imperative to study
interior of relativistic stars in which case these theories could
give different predictions. It is therefore important to study the
situation more closely.

The paper is organized as follows. After giving a short account of
the JBD theory, it has been shown in section 2 that one can find
"Jordan's scalar stars" solutions in isotropic coordinates. The
nature of the "Jordan's scalar stars" with perfect fluid matter
core has been discussed in section 3. Finally the results are
summarized in section 4.

\section{The "Jordan's scalar stars" solutions.}

There is the original spirit \cite{Jordan}, \cite{Brans} of JBD
theory of gravity in which the scalar field $\phi $ is prescribed
to remain strictly massless by forbidding its direct interaction
with matter fields. However, the pure JBD theory can be thought of
as a kind of theory having a non-canonical kinetic term and being
coupled to gravity non-minimally. On the other hand, it is
frequently argued that the spherically symmetric self -
gravitating solitons appear in a number of field systems coupled
to gravity. For example there are boson star solutions in the
Einstein-Klein-Gordon system \cite{Ruffini} and
Einstein-Yang-Mills theory possesses the Bartnik-McKinnon
solutions \cite {Volkov}. It is therefore natural to ask whether a
scalar star solutions might exist in the JBD theory. In fact one
can regard the field equations as being simply the Einstein
equations with a scalar field which interacts with all other
matter fields through the trace of their energy momentum tensor.

The search for the exact spherically symmetric solutions is
continuously of an interest to physicists. These models have been
studied ever since the first solution of Einstein's field equation
was obtained by Schwarzschild. Due to highly non-linear character
of scalar-tensor gravitational theories, a desirable pre-requisite
for studying strong field structure is to have knowledge of exact
explicit solutions of the field equations \cite{Bhadra}. The
Birkhoff's theorem does not hold in the presence of JBD scalar
field, hence several static solutions of the scalar-tensor
theories seem possible in spherically symmetric vacuum case
\cite{OHanlon}. Notice that, it is obvious that we must apply the
Birkhoff theorem even in Einstein theory only in the exterior
vacuum domain outside the star.

 Consider spherically symmetric
spacetime geometry. The most common form of line element of a
D-dimensional spherically symmetric spacetime in comoving
coordinates can be written as

\begin{eqnarray}
ds^2 &=&-g_{tt}\left( r,t\right) dt^2+g_{rr}\left( r,t\right)
dr^2+
\nonumber  \label{eq1} \\
&&+\mathcal{\rho }^2\left( r,t\right) d\Omega _{\left( D-2\right)
}^2. \label{eq1}
\end{eqnarray}

where d$\Omega _{(D-2)}^2$ is the line element on a unit D-2
sphere:

\begin{center}
\begin{eqnarray}
d\Omega _{\left( D-2\right) }^2 = [d\theta _{\left( 0\right) }^2 +
\;\ \;\
\;\ \;\ \;\ \;\ \;\ \;\ \;\ \;\   \label{eq2} \\
\;\ \;\ \;\ \;\ \;\ \;\ \;\ \;\ \;\   \nonumber \\
+
\begin{tabular}{l}
D-3 \\
$\sum $ \\
n=1
\end{tabular}
d\theta _{\left( n\right) }^2\left(
\begin{tabular}{l}
n \\
$\prod $ \\
m=1
\end{tabular}
\sin ^2\theta _{\left( m-1\right) }\right) ].  \nonumber
\end{eqnarray}
\end{center}

One of the basic problems in the description of a source of
gravitational field in relativistic theories is the choice of
proper radial variable \textit{r}. The physical and geometrical
meaning of the radial coordinate \textit{r} is not defined by the
spherical symmetry of the problem and is unknown a priori
\cite{Synge}, \cite{Eddington}.

The forms of static spherically symmetric vacuum solution of the
JBD theories are available in the literature often be explicitly
written down in
isotropic coordinates, defined by $\mathcal{\rho }^2$\textit{(r) = g}$_{rr}$%
\textit{(r) r}$^2$. However, specific solutions, in general, do
not possess the symmetries of the equations they satisfy. The
different gauge may describe different physical solutions of field
equations with the same spherical symmetry \cite{Fiziev}.  As it
was shown in \cite{Fiziev2}, \cite {Aguirregabiria} a nonstandard
gauge fixing for the applications of general theory of relativity
to the stellar physics lead to solutions for some hypothetical
objects with arbitrary large mass, density and size.

Scalar-tensor theories are described by the following action in
the Jordan frame in D-dimensional space-time is:

\begin{eqnarray}
S &=&\int d^Dx\sqrt{-g}(\phi R-\omega \left( \phi \right) g^{\mu
\nu }\nabla
_\mu \phi \nabla _\nu \phi -  \nonumber \\
&&\ \ \;\ \;\ \;\ \;\ \;\ \;\ \;\ \;\ \;\ \;\ \;\ -\lambda \left(
\phi \right) )+S_m.  \label{eq3}
\end{eqnarray}

Here, \textit{R} is the Ricci scalar curvature with respect to the
space-time metric g$_{\mu \nu }$ and S$_m$ denote action of matter
fields. We use units in which gravitational constant \textit{G}=1
and speed of light c=1.
The dynamics of the scalar field $\phi $ depends on the functions $\omega $($%
\phi $) and $\lambda $($\phi $). It should be mentioned that the
different choices of such functions give different scalar-tensor
theories. We restrict
our discussion to the JBD theory which characterized by the functions $%
\lambda $($\phi $) = 0 and $\omega $($\phi $) = $\omega $/$\phi $, where $%
\omega $ is a constant.

Variation of (\ref{eq3}) with respect to g$_{\mu \nu }$ and $\phi
$ gives, respectively, the D-dimensional field equations:

\begin{equation}
R_{\mu \nu }-\frac 12Rg_{\mu \nu }=\frac 1 {\phi \ }T_{\mu \nu
}^M+ T_{\mu \nu }^{JBD},  \label{eq4}
\end{equation}
where

\begin{eqnarray}
T_{\mu \nu }^{JBD} &=& [\frac \omega {\phi ^2}\left( \nabla _\mu
\phi \nabla _\nu \phi -\frac 12g_{\mu \nu }\nabla _\alpha \phi
\nabla
^\alpha \phi \right) +  \nonumber \\[0.01in]
&&\ \ +\frac 1\phi \left( \nabla _\mu \nabla _\nu \phi -g_{\mu \nu
}\nabla _\alpha \nabla ^\alpha \phi \right) ].  \label{eq5}
\end{eqnarray}
and

\begin{equation}
\nabla _\alpha \nabla ^\alpha \phi =\frac{T_\lambda ^{M\ \lambda
}}{\left( D-1\right) +\left( D-2\right) \omega },  \label{eq6}
\end{equation}
and $T_\lambda ^{M\ \lambda }$ is the energy momentum tensor of
ordinary
matter which obeys the conservation equation $T_{\mu \nu ;\lambda }^{M\ }$ g$%
^{\nu \lambda }$= 0.

In this part of article we chose to work in static four
dimensional isotropic spherically symmetric
metric (\ref{eq1}) with $\mathcal{\rho }^2$ = \textit{r}$^2$\textit{g}$_{rr}$%
,
\begin{eqnarray}
ds^2=-e^{2\nu \left( r\right) }dt^2+  e^{2\lambda \left( r\right)
} \left( dr^2+r^2d\Omega^2 \right). \label{eq7}
\end{eqnarray}
because we found them to be the most widespread in literature and
well known Brans solutions written down in this coordinates too.
The use of isotropic coordinates is not a matter of deep principle
and we do not rule out the possibility that there may still be
other representations in other coordinate systems.

Hawking's theorem \cite{Hawking} in JBD states that the only
spherically symmetric solution is static and given (up to
coordinate freedom) by the Schwarzschild metric. However, as we
have seen, even restricting to stationary spherically symmetry JBD
theory has more solutions. Then one consequence of this is the
possibility to use these solutions as interior and match it with
Schwarzschild metric. To get a sense of the nature the static
''Jordan's scalar star'' solutions we consider here the simplest
example, four-dimensional stars with Brans class I  \cite{Brans}
solution as interior

\begin{eqnarray}
\phi &=& \phi _0\left( \frac{1-\frac Br}{1+\frac Br}\right)
^{\frac C A },
  \label{eq21} \\
\lambda &=& \lambda _0+\ln \left[ \left( 1+\frac Br\right) ^2\left( \frac{%
1-\frac Br}{1+\frac Br}\right) ^{\frac{A -C-1}A }\right] ,
\label{eq21} \nonumber \\
\nu &=& \nu _0+\ln \left[ \left( \frac{1-\frac Br}{1+\frac
Br}\right) ^{\frac 1 A }\right] ,  \nonumber
\end{eqnarray}

where:

\begin{eqnarray*}
A =\sqrt{\left( C+1\right) ^2-C\left( 1-\frac{\omega \ C}2\right)
}, \label{2.9}
\end{eqnarray*}

and Schwarzschild solution as exterior of object

\begin{eqnarray}
&& \phi  = 1, \nonumber  \label{eq22} \\
&& \lambda  = ln \left[ \left( 1+\frac \mu r\right) ^2\right] ,
\label{eq22} \\
&& \nu  = \ln \left( \frac{1-\frac \mu r}{1+\frac \mu r}\right),
\nonumber
\end{eqnarray}

In the case of ordinary stars model the discontinuity in the mass
density at the surface entails via the field equations a jump in
second derivations of metric coefficient, but first derivatives
remains continuous so can be used to match to the vacuum solution.
The spherical symmetry by itself automatically implies that once
one calculates the Einstein tensor and goes to an orthonormal
frame. These comments are of course quite standard and in some
form or another implicitly underlie all extant static spherically
symmetric perfect fluid solutions. The same could happen for the
''Jordan's scalar star'' models also.

The JBD scalar field fluid, however, would fail to be a perfect
fluid. That the stress-energy tensor of a scalar stars, unlike a
classical fluid (but the similar as in the case of boson stars
\cite{Schunck} ), is in general
anisotropic. For a spherically symmetric configuration, it becomes diagonal (%
\ref{eq5}), i.e.

T$_{\mu \nu }^{JBD}$($\phi $) = diag ($\varrho $,{\it -p}$_r${\it
, -p}$_{\perp }${\it ,-p}$_{\perp }$).

In contrast to a neutron star, where the ideal fluid approximation
demands the isotropy of the pressure, for spherically symmetric
''Jordan's scalar star'' there are different stresses
\textit{p}$_r$ and \textit{p}$_{\perp }$ in radial or tangential
directions, respectively.

Now in order to justify calling the geometry an exact solution we need an
explicit definition for the constant in Brans solution. The integration
constants of Brans solution \textit{C}, $\lambda _0$, $\nu _0$ and \textit{B}
are arbitrary. However, it is possible to match this solution to the vacuum
Schwarzschild metric. The brief computation yields.

\begin{eqnarray*}
\ B &=& r_{*}\sqrt{\frac{2r_{*}^2-2r_{*}\mu +\mu ^2\left( 2+\omega
\right) }{
-2r_{*}\mu +2\mu ^2+r_{*}^2\left( 2+\omega \right) }} , \ \ \ \ \ \ \ \ \ \ \ \ \ \ \ \nonumber \\
\ C &=& \frac{2\left( r_{*}^2-r_{*}\mu +\mu ^2\right) }{r_{*}\mu \
\omega } , \ \ \ \ \ \ \ \ \ \ \ \ \ \ \ \ \ \ \ \ \ \ \ \ \ \ \ \ \ \ \ \ \nonumber \\
\lambda _0 &=& \ln \left( \frac{\left( 1-\frac{2B}{B+r_{*}}\right)
^{\frac{ 2r_{*}^2+2\mu ^2+r_{*}\mu \left( \omega -2\right) }{A \
r_{*}\mu \ \omega
}}\left( r_{*}+\mu \right) ^2}{r_{*}^2-B^2}\right) ,\ \ \ \ \ \ \ \ \ \nonumber \\
\nu _0 &=& \ln \left( \frac{\left( 1-\frac{2B}{B+r_{*}}\right)
^{-\frac 1{A
\ }}\left( r_{*}-\mu \right) }{r_{*}+\mu }\right) , \ \ \ \ \ \ \ \ \ \ \nonumber \\
\phi _0 &=& \left( \frac{r_{*}+B}{r_{*}-B}\right) ^{\frac{2\left(
r_{*}^2+\mu ^2-r_{*}\mu \right) }{A \ r_{*}\mu \ \omega }} .\ \ \ \ \ \ \ \ \ \ \ \ \ \ \ \ \ \ \ \ \nonumber \\
\end{eqnarray*}
where r$_{*}$ the radius of ''Jordan's scalar star'' where we
match the both solutions and the mass $\mu $ is defined as
Keplerian mass, as seen by a distant observer.

Then in this context, unlike the scalar-tensor theory spirit of
the original JBD gravity, the internal scalar field is not viewed
as a part of the gravitational degrees of freedom but instead is
thought of as playing the role of a matter degree of freedom.
Obviously, this solution has a geometrical topological nature and
may be used in the attempts to reach description of "matter
without matter". Apparently then, these models are also expected
to provide a successful explanation for the phenomena associated
with the dark matter.

Furthermore, there is now a growing consensus that wormholes are
in the same chain of stars and black holes. A remarkable feature
of our model is the fact that, one can represent it as wormhole
the ''bridges'' between two separated Universes of different
natures.

\section{The "Jordan's scalar stars" with perfect fluid matter
core.}

On the other hand, the more difficult task is the construction of
interior perfect fluid solutions which are of great astrophysical
interest. In this line of thought, it is interesting to note that
a relatively new model denoted as a gravastar (gravitational
vacuum star) \cite{Mazur}, consists of a compact object with an
interior de Sitter condensate, governed by an equation of state
given by \textit{p} = - $\varrho $, matched to a shell of finite
thickness with an equation of state \textit{p} = $\varrho $. In
this work, a extension of the gravastar picture is explored by
matching an interior solution with \textit{p}$_M$ = $\varepsilon $
$\varrho$$_M$ to an exterior Schwarzschild solution at a junction
interface, comprising of a "scalar shell". Although this does not
closely describe realistic stars, it can be adequate for
indicating the behavior of mass limits and the stability
properties of equilibrium configurations.

There are some known exact perfect fluid interior solutions in JBD
\cite{Bruckman}, \cite {Kozyrev2},\thinspace \cite{Yazadjiev}.
Those solutions, however, are not physically acceptable: the
pressure is singular at the center or the solutions have not a
well defined boundary. Nevertheless, the exact solutions, even
unrealistic, could qualitatively describe the case of a static,
spherically symmetric perfect fluid ''Jordan's scalar star''.

In what follows we will consider the case of cold
ultrahigh-density static configuration. One can use a
perfect-fluid ordinary matter model with simple equation of state

\begin{eqnarray*}
p_M = \varepsilon \varrho _M.  \label{eq88} \\
\end{eqnarray*}

Because of the considerations above we allow for three different regions
with the three different equations of state,

Interior : 0 $\leq $\textit{\ r} $\ <$\textit{r}$_1$ ,
\textit{p}$_M$ = $\varepsilon $ $\varrho _M$, $\phi $ $\neq $
constant,

Shell : \textit{r}$_1$ $\ <$ \textit{r} $\ <$\textit{r}$_2$ , $\varrho _M$ = \textit{p}$%
_M$ = 0, $\phi $ $\neq $ constant,

Exterior : \textit{r}$_2$ $\ <$ \textit{r} , $\varrho _M$ =
\textit{p}$_M$ = 0, $\phi $ = constant.

At the interfaces \textit{r} = \textit{r}$_1$ and \textit{r} = \textit{r}$_2$%
, we require the scalar field $\phi $ and metric coefficients and
first derivatives of metric coefficients to be continuous,
although the derivatives of $\phi $ be able to discontinuous from
the first order. Since our model is a mixed perfect fluid and
exotic matter core, the requirement that pressure and density
involve the scalar fields tell us that
we can find the surface of the star by locating the first zero of total \textit{p}(%
\textit{r}) or $\varrho $(\textit{r}).

The interior JBD solution for isotropic coordinates can be obtained by using
the method discussed in \cite{Bruckman}. The field equations (\ref{eq4}), (%
\ref{eq6}) can be integrated to give

\begin{eqnarray}
\phi &=& a \ e^{c\ \nu },  \nonumber  \label{eq23} \\
\lambda &=&\lambda _s-\frac{V\ln \left( \left( 1-b \right) r\right) }{b -1}%
,  \nonumber  \label{eq23} \\
\nu &=&\nu_s + Q \ln r,  \label{eq23} \\
\varrho &=&c\ e^{c\ \nu_s -2\lambda _s}r^{c\ Q -2}a \ Q \left(
\left( 1-b \right) r
\right) ^{\frac{2V}{b-1}}\times  \nonumber \\
&&\frac{\left( \left( b -1\right) \left( 1+Q +c \ Q \right)
-V\right) \left( 3+2\omega \right) }{\left( b -1\right) \left(
3\varepsilon -1\right) }.  \nonumber
\end{eqnarray}

where

\begin{eqnarray*}
c &=&\frac{3\varepsilon -1}{\left( 3+2\omega \right) +\left( \omega
+1\right) \left( 3\varepsilon -1\right) }, \\
b &=&1-\frac Q 2+\frac{c \ Q \left( 2+\omega -\varepsilon \left(
3+\omega \right) \right) }{2\left( 3\varepsilon -1\right) }, \\
V &=&\frac{\left( Q -1\right) Q }2+\frac{c \ Q \left( 2+\omega
+ Q \omega -\varepsilon \left( 3 + \omega \right) \right) }{2\left( 3\varepsilon -1\right) }+ \\
&&+c \ Q ^2 \frac{1- \varepsilon \omega +c \ \varepsilon \left(
3+2 \omega  \right) }{2\left( 3\varepsilon -1\right) }, \\
Q &=&\frac{2\varepsilon \left( 2+\omega +3\varepsilon \left(
1+\omega \right) \right) }{\sqrt{2+\omega +6\varepsilon \left(
1+\omega \right)
+9\varepsilon ^2\left( 2+\omega \right) }}\times \\
&&\frac 1{\sqrt{2+\omega +6\varepsilon \left( 1+\omega \right) +\varepsilon
^2\left( 6+\omega \right) }}
\end{eqnarray*}

Consequently possibility of composite models is obtained by
matching the surfaces \textit{r} = \textit{r}$_1$ and \textit{r} =
\textit{r}$_2$ since solutions would be forced to match on both
surfaces. We examine matching across these surfaces between
solutions (\ref{eq23}), (\ref{eq21}) and (\ref {eq22}) to obtain
the values of constants of integration. After some algebra one can
decide

\vspace{6cm}

\begin{eqnarray*}
B &=&\frac{\left( 1-C+ A^2\right) \mu }{2 A }+  \label{2.13} \\
&&+\frac{\sqrt{C^4+\left( A ^2-1\right) ^2-2C^2\left( 1+A ^2\right) }%
\mu }{2A }, \\
C &=&-\frac{2A-\left( b -1\right) \left( 2+Q \right) }{\left( b
-1\right) Q \omega }-\ \ \sqrt{W}
\end{eqnarray*}

where

\begin{eqnarray*}
W &=&\frac{4Q\omega +\left( 2+Q \right) ^2}{Q ^2\omega ^2}+%
\frac{2A^2\left( 2+\omega \right) }{\left( \beta -1\right) ^2Q
^2\omega
^2}- \\
&&-\frac{4A\left( 2+Q +\omega -Q \omega \right) }{\left( b
-1\right) Q ^2\omega ^2}
\end{eqnarray*}

\begin{figure}
\includegraphics {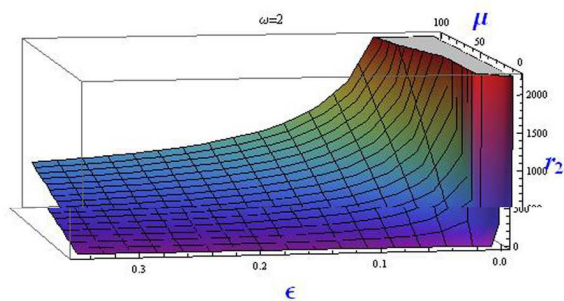}
\caption{The outer radius of the scalar envelope is \textit{r}$_2$
and the total Keplerian  mass inside \textit{r}$_2$ is $\mu $. The
equation of state of the core is \textit{p}$_M$ = $\varepsilon $
$\varrho _M$.}
\end{figure}
%%%%%%%%%%%%%%%%%%%%%%%%%%%%%%%%%%%%%%%%%%%%%%%%%%%%

%\begin{figure}
%\includegraphics {test1.eps}
%\caption{The outer radius of the scalar envelope is \textit{r}$_2$
%and the total Keplerian  mass inside \textit{r}$_2$ is $\mu $. The
%equation of state of the core is \textit{p}$_M$ = $\varepsilon $
%$\varrho _M$.}
%\end{figure}

In a realistic situation, interior matter is present inside a
star. Then, to obtain a viable stellar configuration, one has to
match solutions on both surfaces of scalar shell. Since there are
the freedom of determination integration constant between inside
and outside the star, and hence it is easier to satisfy the
matching condition. Indeed, the matching mechanism has been shown
by us to work in this model. It is clear from this fact that one
can construct viable models of "Jordan's scalar star" with
ordinary perfect fluid equation of state.

The above analysis shows that, for certain values of $\varepsilon
 < 0.3$, the solutions (\ref{eq23}) may actually be interpreted to
yield a gravitation interaction inwardly of objects goes to zero
in the centre and increases beside surfaces. This choice of the
coefficient $\varepsilon $ represents a monotonic increasing
"gravitation constant" in the star interior, and was obtained
previously in the analysis of static spheres  \cite{Bashkov2} as a
specific case of the Newtonian limit of JBD theory.
%%%%%%%%%%%%%%%%%%%%%%%%%%%%%%%%%%%%%%%%%%%%%%%%%%%%
\begin{figure}
\includegraphics {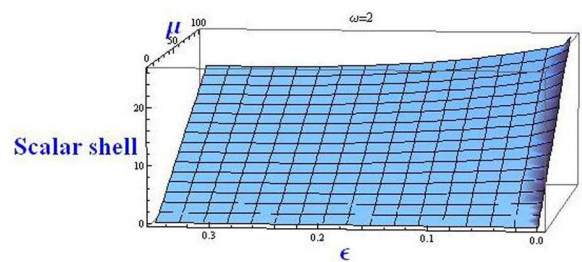}
\caption{Sizes  for shells and Keplerian masses in the "Jordan's
scalar stars".}
\end{figure}
%%%%%%%%%%%%%%%%%%%%%%%%%%%%%%%%%%%%%%%%%%%%%%%%%%%%

\section{Discussion and Conclusions}

Using the key assumption that the Brans class I  \cite{Brans}
solution physically acceptable because the solutions have a well
defined boundary and they can match with the Schwarzschild
exterior solution at the boundary surface it was found the mass
and radius of scalar stars. On the other hand this solution can be
interested in giving a pure field representation of particles.

The new results include matching between exact interior solutions
in the perfect fluid family and the Brans and after that
Schwarzschild solutions. It is clear that the approach to scalar
stellar structure, developed in the present article, calls for
revision of some of widely accepted features of the relativistic
theories of stars. The changes are not based on the critics of
these theories, but on more deep understanding of its
applications, and on attempt to solve some open problems. Hence,
our toy models have an essential impact only on the theory of the
interior of relativistic stars, and on theory of spreading of
different physical fields in stars, and around the stars. Among
others, this solution one can assume as wormhole the ''bridges''
between separated Schwarzschild and Brans Universes.

In this context, it is demonstrated that our toy model can
successfully predict the emergence of dark matter in terms of a
self-gravitating spacetime solution to the JBD field equations
with non-trivial energy density of the JBD scalar field which was
absent in the context of general relativity where the Newton's
constant is strictly a constant having no dynamics. Bearing the
above evaluation in mind, let us comment on the other specific
models of scalar and vector metric gravity. The some vector-metric
theories one shared the similar structure of energy momentum
tensor \cite {Will2}. Therefore, we expect that the same problem
arises in these theories. We have shown that this predicts a
rather interesting physics for the range from stars to clusters of
galaxies. First of all we point out that in this model the stars
acquires features of a two-component objects (ordinary matter and
scalar or vector field) whose distribution in the observed
3-dimensional volume can has, in an addition to standard model, an
envelope of scalar or vector fields. Moreover, such a picture can
represents a Schwarzschild background, while the interior should
be considered as vacuum solution of scalar or vector-metric
theories which defined a Keplerian mass of this object.

Some final remarks, now the local value of the Newtonian
"gravitation constant" measured only near the Earth. For central
and peripheral parts of Galaxy value of "gravitation constant" can
be vastly differ from Newtonian value. In this point of view the
dark matter problem may be explain by a mixture of various
interacting scalar and vector field potentials inside the galaxies
and galaxies clusters.

\section{Acknowledgments}

 The author is grateful to R.A. Daishev and S.V. Sushkov for the useful
discussions. The work was supported in part by the Institute of
Applied Problems.

\section{Appendix }

The possibility that spacetime has more than four dimensions, was
first contemplated by Nordstr\"om \cite{Nordstrom}. It is helpful
here to emphasize the equivalence between the (D + 1) Kaluza-Klein
theories with empty D-dimensional JBD theories when $\omega $ = 0
\cite{Rippl}. Now the idea that spacetime has extra dimensions
lies today at the heart of the most theories of unification of the
fundamental interactions present in Nature, but the introduction
of the fifth and higher dimensions requires a careful approach. In
particular, static spherically symmetric vacuum solutions in
D-dimensional scalar-tensor theories shed new light on the complex
features of objects in these models.

 According to the standard
textbooks \cite{Wehus} the static spherically symmetric vacuum JBD
solutions can often not be explicitly written down in standard
Hilbert or curvature coordinates and it is better to work in
isotropic coordinates. However, the first exact solution of JBD
field equations in widespread Hilbert coordinates were obtained in
parametric form by Heckmann \cite{Heckmann}, soon after Jordan
proposed scalar-tensor theory. This solution describes the
geometry of the space-time exterior to a prefect fluid sphere in
hydrostatic equilibrium. The change of variables technique for
obtaining the similar static solutions in D dimensions are known
by now \cite{Kozyrev}. Choose the static spherically symmetric
metric (\ref{eq1}) in Hilbert gauge: $\mathcal{\rho }$=$\hat{\
r}$, g$_{rr}$ = e$^{2\lambda \left( \hat{r}\right) }$ and g$_{tt}$
=e$^{2\nu \left( \hat{r}\right) }$,

\begin{eqnarray}
ds^2=-e^{2\nu \left( \hat{r}\right) }dt^2+e^{2\lambda \left(
\hat{r}\right) }d \hat{r}^2+ \hat{r}^2d\Omega^2.  \label{eq7}
\end{eqnarray}

It is actually possible to change variables so that one will
replace variable $\hat{\ r}$ by $\hat{\ r}$($\nu $) then the field
equations (\ref{eq4}), (\ref{eq6}) take a form:

\begin{eqnarray}
  -1+\lambda ^{\prime }-\frac{\left( D-2\right) \hat{r}^{\prime }}{ \hat{r}  }+\frac{\hat{r}^{\prime \prime }}
{ \hat{r} ^{\prime }} = \frac{\phi ^{\prime \prime }}{\phi ^{\prime }},\label{eq8}\\
 -1+\lambda ^{\prime }-\frac{\left( D-2\right)\hat{r}^{\prime}\lambda ^{\prime }}{ \hat{r}  }
+\frac{\hat{r}^{\prime \prime }}{  \hat{r} ^{\prime }}=  \label{eq9}\\[0.01in]
 - \frac{\phi ^{\prime }\lambda^{\prime }}\phi + \frac{\omega
\phi ^{\prime 2}}{\phi ^2}-\frac{\phi ^{\prime }\hat{r}^{\prime
\prime }}{\phi \hat{r} ^{\prime }}+\frac{\phi ^{\prime \prime
}}{\phi}, \nonumber \\
 -1+\lambda ^{\prime }-\frac{\left( D-3\right) \left( -1+e^{2\lambda }\right)\hat{r}^{\prime }}{ \hat{r}  } =
\frac{\phi \prime }\phi , \label{eq10} \\
-1+\lambda^{\prime}-\frac{\left(D-2\right)\hat{r}^{\prime}}{
\hat{r}  } + \frac{\hat{r}^{\prime \prime }} { \hat{r} ^{\prime }}
= \frac{\phi \prime }\phi. \label{eq11}
\end{eqnarray}

where now $\nu $ is a new variable and the primes denote
derivatives with respect to $\nu $.

By eliminating $\hat{\ r}$($\nu $) and $\lambda $($\nu $) from
equations (\ref{eq8}) and (\ref{eq11}) we can obtain the following
equation

\begin{eqnarray}
&& \ \frac{\phi ^{\prime \prime }}{\phi ^{\prime }}-\frac{\phi
\prime }\phi =0. \label{eq12}
\end{eqnarray}

Eq. (\ref{eq12}) can immediately be integrated to give

\begin{eqnarray}
&& \ \phi = \phi _0 \ e^{\nu b }.  \label{eq13}
\end{eqnarray}
where $\phi _0$ and \textit{b} are an arbitrary constants of
integration. Specifically, if the scalar field is constant ($\phi
$ = const) then the solution of Eqs. (\ref{eq8}) - (\ref{eq11}) is
a D - dimensional Schwarzschild solution.

\begin{eqnarray}
\lambda &=&-\nu ,  \label{eq14} \\
\hat{r} &=&\left( -1+e^{2\nu }\right) ^{\frac 1{3-D}}.  \nonumber
\end{eqnarray}

After a straightforward calculation using Eqs. (\ref{eq8}) -
(\ref{eq11}) and (\ref{eq13}), we obtain the three possible
solutions for metric components and function $\hat{\ r}$($\nu $) :

\begin{flushleft}
\begin{eqnarray}
\lambda  = \ln \left( \sqrt{\frac A{A+\left( 1+b \right)
^2}}\sec \left[ \sqrt{A}\left( q +\nu \right) \right]\right), \label{eq15} \\
\hat{r} = a \{\left( D-3\right) e^{\left( 1+b \right)
\nu}(\left( 1+ b \right)\times  \ \ \ \ \ \ \ \ \ \ \ \ \ \ \ \nonumber \\
\cos \left[ \sqrt{A}\left( q + \nu \right) \right] +\sin \left[
\sqrt{A}\left( \gamma +\nu \right) \right] )\}^{\frac 1{D-3}}
\nonumber
\end{eqnarray}
\end{flushleft}

\begin{eqnarray}
\lambda = \ln \left( \sqrt{\frac A{A+\left( 1+b \right) ^2}}\sec
\left[
\sqrt{A}\left( q +\nu \right) \right]\right),  \label{eq16} \\
\hat{r} = a e^{-\frac{\left( 1+b \right)\left( \gamma +\nu \right)
+ arccoth\left[ \frac{\left( 1+b \right) \cot \left[
\sqrt{A}\left( q
+\nu \right) \right] }{\sqrt{A}}\right] }{D-3}} \times  \nonumber \\
\{\left( D-2\right) \left( A-\left( 1+b \right) ^2\right)
- \nonumber \\
- \left( A+\left( 1+b \right) ^2\right) \cos \left[
2\sqrt{A}\left( q +\nu \right) \right] \}^{\frac 1{2D-5}}
\nonumber
\end{eqnarray}

\begin{eqnarray}
\lambda  &=&\ln \left( 2\sqrt{-A\ a } \ e^{\nu \sqrt{-A}}\right) -
\label{eq17} \\
&&-\ln (\left( 1+b \right) -\sqrt{-A}-  \nonumber \\
&&-e^{2\nu \sqrt{-A}}\left( \left( 1+b \right) +\sqrt{-A}\right) a
),
\nonumber \\
\hat{r} &=& a \left( \frac{\sqrt{-A\ }}{\left( D-3\right)
e^{\left( \left(
1+b \right) -\sqrt{-A}\right) \nu }}\right) ^{\frac 1{D-3}} \times \nonumber \\
&&\left( \frac 2{-1+q \ e^{2\nu \sqrt{-A}}}\right) ^{\frac
1{D-3}}, \nonumber
\end{eqnarray}
where \textit{q} and \textit{a} are an arbitrary constants of
integration and

\[
A=\frac{\left( b -1\right) ^2\left( 1+2\omega \right) -\left(
D-1\right) \left( 1+b ^2\left( 1+\omega \right) \right) }{D-2}
\]

The derivation extends previous results of Heckmann for higher
dimensions. By choosing some values of arbitrary constant
\textit{b} several special classes of static vacuum solutions can
be obtained in the framework of the scalar-tensor model. Assuming
that the arbitrary constant \textit{b}  = -1, one can obtain
another solution:

\begin{eqnarray}
\phi &=&\phi _0 \ e^{-\nu },\ \ \ \ \ \ \ \ \ \ \ \ \ \ \ \ \ \ \
\ \ \ \ \ \ \ \ \ \ \ \ \ \ \ \ \ \ \label{eq18}
\end{eqnarray}
\begin{eqnarray}
\lambda = \frac{1}{2} \ln \left[- \sinh \left( \frac{\nu
\sqrt{\left( \omega+2
\right)\left( D-3 \right)}}{\sqrt{D-2}}- q \right)^{-2}\right] ,  \nonumber  \\
\hat{r} = a \cosh \left[ q -\frac{\nu \sqrt{\left( \omega+2
\right)\left( D-3 \right)}}{\sqrt{D-2}} \right]^{\frac{1}{3-D}}\ \
\ \ \ \ \ \ \ \ \ \nonumber
\end{eqnarray}

Alternatively, instead of taking the \textit{b} = - 1, we can form
the constant \textit{b} = 2/$\omega $ and obtain an different
description of the solution.
\begin{eqnarray}
\phi &=&\phi _0 \ e^{\frac{2\nu }\omega \ },  \nonumber  \label{eq19} \\
\lambda &=&q -\frac{ \nu \left( 2+\omega \right) }\omega
\label{eq19} \\
\hat{r} &=&a \left(- e^{2q}+e^{\frac{2 \nu \left( 2+\omega \right)
}\omega}\right) ^{-\frac 1{D-3}} \nonumber
\end{eqnarray}

Now using the constant from the equation (\ref{eq13}), viz.

\[
b =\pm \frac{1-\sqrt{\left( 3-D \right) \left( \omega \left(
D-2\right)+ D - 1 \right) }}{D-2+\omega \left( D-3\right) }
\]

we get

\begin{eqnarray}
\phi &=&\phi _0 \ e^{\nu b \ },  \nonumber \\
\lambda &=&-\ln \left[ B \nu + q \right]
\label{eq20} \\
\hat{r} &=&a \left( e^{B \nu }\left( -1+q +B \nu \right) \right)
^{\frac 1{3-D}}  \nonumber
\end{eqnarray}

So far the solutions found is a simple mathematically consistent
solutions. It was constructed to clarify the method described in
this section. In order to obtain a physically acceptable solution,
it is necessary to carry out a more careful analysis.

We have explicitly characterized the D - dimensional spacetime
metrics corresponding to the JBD static spherically symmetric
geometries in a relatively straightforward manner. Although a
tremendous amount is already known concerning static spherically
symmetric spacetimes the particular approach adopted in the
present article may be useful for understanding the inherent
non-linear character of JBD gravitational theory.

While the vacuum JBD field equations possess a well-known D -
dimensional Schwarzschild solution for an isolated mass
\textit{M}, one can transform other above vacuum spherically
symmetric solutions into an interior of "Jordan's scalar star". If
we consider models in which the central region contains the scalar
field only, then it can be used to study the interior structure of
the D - dimensional relativistic objects with anisotropic pressure
of exotic matter.

\renewcommand{\refname}{References}


\begin{thebibliography}{99}
\bibitem{Mach}  {\small E. Mach, The Science of Mechanics (Open Court,
LaSalle, IL, 1960). Mach's Principle: From Newton's Bucket to
Quantum Gravity, edited by J. Barbour and H. Pfister (Birkhauser,
Boston, MA, 1995). }

\bibitem{Dirac}  {\small Dirac, P. A. M., Long range forces and broken
symmetries, Proc. R. Soc. Lond. A333, 403 (1973) }

\bibitem{Visser}  {\small M. Visser, Lorentzian wormholes: from Einstein to
Hawking, Springer-Verlag, New York, Inc. (1996). }

\bibitem{Nandi}  {\small K. K. Nandi, B. Bhattacharjee, S. M. K. Alam, J.
Evans, Brans-Dicke wormholes in the Jordan and Einstein frames, Phys. Rev. D
57, 823 - 828 (1998); A. Bhadra, K. Sarkar, Wormholes in vacuum Brans-Dicke
theory, arXiv:gr-qc/0503004, (2005). }

\bibitem{Jordan}  {\small P. Jordan, Schwerkraft und Weltall, Vieweg
(Braunschweig) 1955. }

\bibitem{Brans}  {\small C. Brans and R. H. Dicke, Mach's Principle and a
Relativistic Theory of Gravitation, Phys. Rev. 124, 925-935, (1961). }

\bibitem{Matsuda}  {\small T. Matsuda, On the Gravitational Collapse in
Brans-Dicke Theory of Gravity, Progr. Theor. Phys. 47, 738-740, (1972). }

\bibitem{Romero}  {\small C. Romero and A. Barros, Brans-Dicke vacuum
solutions and the cosmological constant: A qualitative analysis, Gen. Rel.
Grav. 25, 491-502 (1993). }

\bibitem{Scheel}  {\small M.A. Scheel, S.L. Shapiro and S.A. Teukolsky,
Collapse to black holes in Brans-Dicke theory. II. Comparison with general
relativity, Phys. Rev. D 51, 4236 - 4249 (1995). }

\bibitem{Will}  {\small C. M. Will, The Confrontation between General
Relativity and Experiment, Living Rev. Relativity 9 (2006), http://
www.livingreviews.org/lrr-2006-3 }

\bibitem{Fujii}  {\small Y. Fujii, K. Maeda, The ScalarTensor Theory of
Gravitation, Cambridge University Press, (2003). }

\bibitem{Eubanks}  {\small T. M. Eubanks et al. (1999). Advances in solar
system tests of gravity. In: Proc. of The Joint APS/AAPT 1997
Meeting, 18-21 April 1997, Washington D.C. BAAS, Online preprint,
ftp://casa.usno.navy.mil/navnet/postscript/\\ prd\_15.ps. }

\bibitem{Hawking}  {\small S. W. Hawking, Black holes in the Brans-Dicke
Theory of gravitation, Commun.Math. Phys. 25, 167171 (1972). }

\bibitem{Johnson}  {\small M. Johnson, Lett. Nuovo Cimento 4, 323327 (1972) }

\bibitem{Thorne}  {\small K.S. Thorne, J.J.Dykla, Black Holes in the
Dicke-Brans-Jordan Theory of Gravity, Ap. J. 166, L35L38 (1971) }

\bibitem{Will2}  {\small C. M. Will, Theory of experiment in Gravitational
Physics (Camb. Univ. Press, Cambridge, 1993). }

\bibitem{Bashkov}  {\small V. Bashkov, S. Kozyrev, Problems of high energy
physics and field theory, 22, Protvino, (1991). }

\bibitem{Bhadra}  {\small A. Bhadra, K. Sarkar, On static spherically
symmetric solutions of the vacuum Brans-Dicke theory arXiv:gr-qc/0505141
(2005). }

\bibitem{OHanlon}  {\small J. O'Hanlon, B. O. J. Tupper, Vacuum-field
solutions in the brans-dicke theory, Il Nuovo Cimento B (1971-1996), V. 7,
N. 2, 305-312, (1972). }

\bibitem{Synge}  {\small J. L. Synge, Relativity: The General Theory, North
Holland Publ. Comp., Amsterdam, 1960. }

\bibitem{Eddington}  {\small A. S. Eddington, The mathematical theory of
relativity, 2nd ed. Cambridge, University Press, 1930 (repr.1963). }

\bibitem{Fiziev}  {\small P. P. Fiziev, Gravitational Field of Massive Point
Particle in General Relativity, arXiv :gr-qc/0306088 (2003). }

\bibitem{Fiziev2}  {\small P.P.Fiziev, Novel Geometrical Models of
Relativistic Stars. arXiv :astro-ph/0409456 (2004) }

\bibitem{Aguirregabiria}  {\small J. M. Aguirregabiria, Ll. Bel, Extreme
objects with arbitrary large mass, or density, and arbitrary size arXiv
:gr-qc/0105043, Gen. Rel. and Grav. 33, 2049 (2001). }

\bibitem{Nordstrom}  {\small G. Nordstr\"om, ''\"Uber die M\"oglichkeit das
elektromagnetische Feld und das Gravitationsfeld zu vereinigen,'' Phys.
Zeit. 15, 504506 (1914). }

\bibitem{Rippl}  {\small S. Rippl, C. Romero, R. Tavakol1, D-Dimensional
Gravity from (D + 1) Dimensions, gr-qc/9511016,(1995) }

\bibitem{Wehus}  {\small I. K. Wehus, F. Ravndal (2006). Gravity coupled to
a scalar field in extra dimensions. Preprint arXiv:gr-qc/0610048v2 }

\bibitem{Heckmann}  {\small O. Heckmann, P. Jordan, R.Fricke, Astroph., Zur
erweiterten Gravitationstheorie, Z. 28, 113-149, (1951). }

\bibitem{Kozyrev}  {\small S. Kozyrev, A D-dimensional Heckmann-like
solution of Jordan-Brans-Dicke theory, arXiv: 0712.2894v1 [gr-qc],(2007) }

\bibitem{Ruffini}  {\small R. Ruffini and S. Bonazzola, ''Systems of
Self-Gravitating Particles in General Relativity and the Concept of an
Equation of State''. Phys. Rev. 187: 1767-1783, (1969). }

\bibitem{Volkov}  {\small S. Volkov, D. Gal'tsov, Gravitating Non-Abelian
Solitons and Black Holes with Yang-Mills Fields, Physics Reports, 319, 1,
1-83, (1999); hep-th/9810070 (1998). }

\bibitem{Schunck}  {\small F. E. Schunck E. W. Mielke, Topical review,
General relativistic boson stars, Class. Quantum Grav. 20 R301-R356 (2003),
arXiv:0801.0307 [astro-ph], (2008). }

\bibitem{Mazur}  {\small P. O. Mazur and E. Mottola, "Gravitational
Condensate Stars: An Alternative to Black Holes,"
[arXiv:gr-qc/0109035]; }

\bibitem{Bruckman}  {\small W. Bruckman, E. Kazes, Properties of the
solutions of cold ultradense configurations in the Brans-Dicke theory, Phys.
Rev. D 16, 261-268 (1977) }

\bibitem{Kozyrev2}  {\small S. Kozyrev, Properties of the static,
spherically symmetric solutions in the Jordan-Brans-Dicke theory,
gr-qc/0207039 }

\bibitem{Yazadjiev}  {\small S. S. Yazadjiev, Interior perfect fluid
scalar-tensor solution, gr-qc/0312019 (2003) }

\bibitem{Bashkov2}  {\small V.I.Bashkov, S.M.Kozyrev, Dark matter an effect of gravitation
permeability of material in Jordan, Brance - Dicke theory, arXiv
:gr-qc/0103009 (2001). }
\end{thebibliography}
\end{document}